\documentclass[preprint,preprintnumbers,amsmath,amssymb, floatfix]{revtex4}
\usepackage{bm}
\def\be{\begin{equation}}
\def\ee{\end{equation}}
\def\beq{\begin{eqnarray}}
\def\eeq{\end{eqnarray}}

\begin{document} \openup8pt 
\preprint{smw-let-01-06}
\title{Universal Theory of Relativity and the ``Unification''
of Fundamental Physical Interactions}

\author{Sanjay M Wagh}

\affiliation{Central India Research Institute, Post Box 606,
Laxminagar, Nagpur 440 022, India \vspace{1.5in}}

\begin{abstract}
The ``unification'' of fundamental physical forces (interactions)
imagines a ``single'' conceptual entity using which {\em all\/}
the observable or physical phenomena, {\em ie}, changes to
physical bodies, would be suitably describable. The physical,
conceptual and mathematical, framework which achieves this is that
of the recently proposed Universal Theory of Relativity
\cite{smw-utr}. Here, we argue that the mathematical framework
required to achieve the ``unification'' should be that of the
general Category Theory. There are certain unanswered mathematical
questions arising out of this context. In the sequel, we also
point out these issues for the wider attention.
\end{abstract}

\email{cirinag_ngp@sancharnet.in}

\date{February 7, 2006}
\maketitle

\newpage
Physics is our attempt to conceptually grasp the happenings of the
observable world. Various physical concepts are also succinctly
expressible in the language of mathematics. Laws of Physics are
therefore mathematical statements about mathematical structures
representing ``observable'' or ``physical'' bodies. Unless the
theory (concepts and their mathematical representations, both) is
appropriate, it will fail to explain, at least, ``some''
observations. Needs of ``appropriate'' physical concepts and
mathematical structures to represent them are, both, then evident
\footnote{As examples, we recall the formulation of the concept of
motion using velocity and its mathematical representation
initially as a scalar and later as a vector quantity.}.

Experiments ``verify'' theoretical explanations of phenomena and,
in turn, indicate the appropriateness of our choice of, both, the
physical conceptions and the mathematical structures representing
them. There also are purely logical methods to decide, at least
partly, this appropriateness. These methods determine the {\em
mutual compatibility\/} of our concepts, {\em ie}, the {\em
internal consistency\/} of the theoretical framework.

Using any of these above methods, we then judiciously accept or
reject any conceptual framework as an admissible theory of the
observable world. When an internally consistent theory fails to
explain some observations, we need to {\em expand\/} the
conceptual basis of that theory and, hence, mathematical
structures representing those concepts. As an acceptable
explanation of the observable world, the conceptual framework of
the ``expanded'' theory must also be internally consistent in the
sense of Logic.

Currently, physical explanations rely on four basic forces (of
newtonian conceptual origin as ``means'' to cause ``changes'' to
physical bodies), {\em viz}, gravity, electromagnetism, weak
nuclear, and strong nuclear interactions. The present goal of
theoretical physics is of ``unifying'' these four independent
forces. Clearly, this aim requires appropriate physical concepts
and their mathematical representations, both.

Einstein \cite{ein-pop, schlipp}, from (only an apparently)
different perspective than that of the above, arrived at the
Principle of General Relativity that the Laws of Physics be
applicable with respect to \underline{all} the systems of
reference, in relative acceleration or not, {\em without unnatural
forces} (whose origin is not in physical or observable bodies)
{\em entering into them}. Then, the Laws of Physics should be
based on the same mathematical structures, and be also the same
mathematical statements, for all the reference systems.

Changes in physical bodies are the {\em physical phenomena}.
Physical reference systems are physical bodies used as reference.
This situation provides \cite{smw-utr, smw-sars, smw-sams,smw-ppg}
us a {\em guiding principle}: mathematical structure(s)
representing physical bodies be such that phenomena become
`changes' to (mathematical structure of) reference systems
themselves.

Here, the unification of basic forces is closely related to
Einstein's principle of general relativity and the aforementioned
guiding principle, the former principle helping us with the
formulations of various physical laws and the latter principle
helping us select the unifying mathematical structure. This is the
theoretical framework which I had called as the {\em Universal
Theory of Relativity\/} \cite{smw-utr}.

Now, the ``unification'' of fundamental physical interactions must
postulate ``some'' {\em single\/} mathematical entity that
``represents'' not only {\em all\/} the characteristics of
physical bodies but also their ``changes'' (mathematical
transformations).

Then, we note that a mathematical transformation essentially
``knows'' about the mathematical structure it ``transforms''. This
single concept, that of the transformation of a mathematical
structure representing {\em all\/} the characteristics of physical
bodies, appears to possess therefore the ingredients necessary to
be the single conceptual entity capable of ``unifying'' the four
fundamental forces.

This is the {\em same\/} as in the case of a category. A {\em
category}, $\mathcal{C}$, is usually defined in terms of {\em
two\/} collections - the first, $\mathcal{C}_o$, of {\em
objects\/} and the second, $\mathcal{C}_A$, of {\em arrows or
transformations\/} and {\em four suitable operations\/} on
collections $\mathcal{C}_o$ and $\mathcal{C}_A$ satisfying
conditions naturally arising for them to be mutually compatible
operations.

However, a category is also definable \cite{Law-66, Mac71,
bremen-acc, macmor} in terms of only $\mathcal{C}_A$, in an
object-free manner. Objects serve only to index (identity) arrows
as far as the functions \footnote{A function $f$ with domain $X$
and codomain $Y$ is treated here as a triple $(X,f,Y)$, where
$f\subseteq X\times Y$ is a relation such that for each $x\in X$,
there exists a unique $y\in Y$ with $(x,y)\in f$. We also write
$y=f(x)$ or $x\mapsto f(x)$.} from a subset of
$\mathcal{C}_A\times\, \mathcal{C}_A$ to $\mathcal{C}_A$ are
concerned. Thus, a category can be defined using only
$\mathcal{C}_A$ and the binary operation of {\em composition\/} of
arrows, an operation which is not always defined and is subject to
naturally arising compatibility conditions.

Categorical foundations of the aim of ``unification'' are then
manifest. We focus either on a suitable mathematical structure
possessing {\em all\/} the characteristics of physical bodies and
consider appropriate means (equations) arising out of its possible
transformations to arrive at physically verifiable conclusions or,
completely equivalently, on {\em general\/} transformations (as
arrows of an abstract category) and `extract' {\em all\/} the
characteristics of physical bodies  from these transformations
alone.

We now turn to some reasons as to why only the most general
mathematical framework of the Category Theory, and nothing short
of it, is suitable for concepts behind universal relativity or the
unification of basic interactions.

The representation (originating with Euclidean and Cartesian
conceptions) of physical bodies by points is inadequate vis-a'-vis
the principle of general relativity and the aforementioned guiding
principle, both: with it, reference systems do not ``change'' with
the occurrence of observable phenomena. We thus require a
mathematical notion other than a point to represent physical
bodies.

We may then choose to represent a physical body by a collection of
``points''. Now, any physical body can be considered to possess a
``boundary''. (But, see discussion later.) Then, the ``interior''
of such a set with boundary is an ``open'' subset of suitable
topological space $X$ constructible out of many such points. Thus,
we could represent physical bodies by open subsets of the
corresponding topological space $X$.

However, if we consider a physical body as a (collection of) open
subset(s) of an underlying topological space $X$, then it will,
mathematically equivalently, be also represented \cite{smw-sars,
smw-sams, smw-ppg} as a {\em spatial Frame} \cite{banaschewski}.

Technically speaking, a {\em Frame\/} is a complete lattice
\footnote{A {\em lattice}, $L$, is a partially ordered set (poset)
such that for all $x, y \in L$ there exists a lowest upper bound
(lub or sup or join), denoted as $x\vee y$, and a greatest lower
bound (glb or inf or meet), denoted as $x \wedge y$. A poset $P$
is {\em complete\/} iff every subset of $P$ has a lowest upper
bound and a greatest lower bound. If there exist elements $0,
\mathfrak{1} \in L$ such that $0\leq x\leq \mathfrak{1}$ for all
$x\in L$, we call $L$ a {\em complete lattice\/} and it is a
complete poset. [For sets, $\wedge$ corresponds to
set-intersection and $\vee$ corresponds to set-union.]}, denoted
by $L$, in which  binary meet distributes over arbitrary joins,
{\em ie}, $ x\wedge \bigvee S = \bigvee\left\{ x\wedge y|y\in
S\right\}$ for all $x\in L$ and $S\subseteq L$. In considering a
{\em Frame}, we thus focus essentially on the concept of an
``order'' definable on a set, for example, order by ``inclusion''
of one open subset in another open subset. (This is in contrast to
the notion of a neighborhood of a point on which topological
considerations focus by generalizing relevant properties of the
(Euclidean) real line.) In this last case, we regain the topology
on a set from the order by inclusion on the class of its (open)
subsets, and such {\em Frames\/} are called the {\em spatial
Frames}.

For a space $X$, the lattice of its open subsets, ${\cal O}X$,
forms, trivially, a {\em spatial Frame}. Now, the topology of a
space $X$ is always obtainable from suitable {\em spatial Frame},
but the {\em Frame}, in general, need not \footnote{Not every
``order'' definable on a set is `equivalent' to the order by
inclusion on any topology-forming collection of its (open)
subsets. That is why every {\em Frame\/} is {\em not spatial}.}
correspond to topology on any space $X$.

In categorical notions, the category $\mathbb{S}p\mathbb{F}rm$ of
spatial {\em Frames\/} is a full sub-category of the category
$\mathbb{F}rm$ of {\em Frames}. Also, the category opposite
\footnote{Category Opposite or Dual to a category $\mathcal{C}$
has the same objects as $\mathcal{C}$ but directions of all the
arrows as well as the orders of all the compositions of arrows in
$\mathcal{C}$ are {\em reversed}.} to $\mathbb{F}rm$ is called the
category $\mathbb{L}oc$ of {\em locales}, the notion of a
``Locale'' being reminiscent of that of the neighborhood of a
point of a set. This is the framework of the Point-free Topology
\cite{banaschewski}.

Now, the concept of the boundary of a physical body is not any
fundamental physical notion of the collection of points making up
that body. We could, in its place, then focus on ``orders''
imposable on the set of points constituting a physical body, that
is to say, we could choose to represent a physical body by a {\em
Frame}.

Then, representation of some properties of material bodies will
involve {\em mathematical structures\/}  (measures or
generalizations thereof) {\em defined over Frames or over the
arrows of\/} $\mathbb{F}rm$. Corresponding notions have not been
developed as yet.

A map $h:L\to M$ between {\em Frames\/} $L$ and $M$ preserving
finite meets (including $\mathfrak{1}$) and arbitrary joins
(including $0$) is called a {\em Frame homomorphism}. Such maps
could then represent observable phenomena affecting physical
reference systems.

So far, the collection of points constituting a physical body is
treated as a set from the intuitive set theory (to avoid paradoxes
like Russell's paradox). This too is not mandatory. The
set-theoretic restrictions can then be relaxed \cite{Mac71,
bremen-acc} in the setting of a {\em Quasi-Category\/} which uses
a collection of objects (classes) as a conglomerate \footnote{A
{\em class\/} is a collection of sets: for any property $P$, we
can form a class of all sets with property $P$. But, there is no
surjection from a set to a class that itself is not a set. Every
set is a class. A {\em conglomerate\/} is a collection of classes:
for any property $P$, we can form a conglomerate of classes with
property $P$. Moreover, we assume an Axiom of Choice for
Conglomerates: for each surjection $f:X\to Y$ of conglomerates,
there exists an injection $g:Y\to X$ with $f\circ g = id_Y$, the
identity of $Y$. Every class is also a conglomerate. Usual
set-theoretic constructions are permissible for classes and
conglomerates.}.

We could then represent a physical body as an object (class) of
the Quasi-Category. Presently however, a general mathematical
structure (measures or generalizations thereof) to represent {\em
all\/} the physical characteristics of observable bodies does not
seem to have been developed within this framework or within its
object-free form.

Nonetheless, Category Theory \cite{Law-66, Mac71, bremen-acc,
macmor} does appear to provide \cite{smw-sars, smw-sams, smw-ppg}
a mathematical basis to conceptions behind the unification of all
the basic physical interactions.

Notably, very general physical and mathematical conceptions are
identical then. Much work is of course needed before the physical
concepts of universal relativity or of unification of forces can
be mathematically represented in a satisfactory way.

In the end, we emphasize that the approach adopted here is also
{\em logically compelling\/} if Mathematics is indeed the tool to
represent our physical conceptions.

\acknowledgments I am grateful to Partha Ghosh, Gareth Amery,
Sunil Maharaj and many others for helpful discussions. I am also
indebted to Hemant Wagh, MD (Psychiatry) and Dimi Chakalov
(Psychologist) for raising certain conceptual issues and for
related discussions. This work is dedicated to the memory of Prof
A K Raychaudhury.

\end{document}